# Protein - Small Molecule docking with receptor flexibility in iMOLSDOCK

D. Sam Paul and N. Gautham *

Centre of Advanced Study in Crystallography and Biophysics,

University of Madras, Chennai 600025, India.

First Author              :  Dr. D. Sam Paul

                                 Centre of Advanced Study in Crystallography and Biophysics

                                 Guindy Campus, Chennai 600025, India.

                                 ORCID :  0000-0002-8272-9561

Corresponding Author:  Dr. N. Gautham, Professor Emeritus,

                                 Centre of Advanced Study in Crystallography and Biophysics,

                                 Guindy Campus, Chennai 600025, India.

**Acknowledgements**

We thank the Department of Science and Technology, Government of India, for financial support.

We also thank the University Grants Commission for support under the CAS program.

**Conflict of interest**: The authors declare no conflict of interest.

**Abstract:**

We have earlier reported the iMOLSDOCK technique to perform 'induced-fit' peptide-protein docking. iMOLSDOCK uses the mutually orthogonal Latin squares (MOLS) technique to sample the conformation and the docking pose of the small molecule ligand and also the flexible residues of the receptor protein, and arrive at the optimum pose and conformation. In this paper we report the extension carried out in iMOLSDOCK to dock nonpeptide small molecule ligands to receptor proteins. We have benchmarked and validated iMOLSDOCK with a dataset of 34 protein-ligand complexes with nonpeptide small molecules as ligands. We have also compared iMOLSDOCK with other flexible receptor docking tools GOLD v5.2.1 and AutoDock Vina. The results obtained show that the method works better than these two algorithms, though it consumes more computer time. The source code and binary of MOLS 2.0 (under a GNU Lesser General Public License) are freely available for download at https://sourceforge.net/projects/mols2-0/files/

**Keywords:**

Molecular docking, protein-ligand docking, induced-fit docking, side-chain flexibility, iMOLSDOCK, mutually orthogonal Latin squares.

**Introduction**

Molecular docking programs are used to find the best binding pose of a ligand in the protein binding site. Apart from protein-ligand interactions, they are also used to study protein-protein and protein-DNA interactions. In the past 30 years, a large number of docking programs have been developed using various techniques [1]. Docking problems are computationally addressed by combining an accurate representation of the intermolecular interactions, an efficient algorithm and a scoring function. The algorithm in the docking program searches for potential binding modes and the scoring function discriminates and ranks the predicted structures [2].

Initial docking methods treated the ligand and receptor both as fixed structures. The development progressed by including variations in the internal degrees of freedom of the ligand [3]. However, receptor proteins do not remain rigid, but change conformation to accommodate the ligand [4–6]. Active site plasticity of the receptor proteins was also established by comparing the apo and holo forms of proteins. Though main-chain conformations are largely preserved, in most of the proteins significant differences in side-chain conformations occur upon ligand binding [7, 8]. Therefore, docking algorithms which allowed receptor flexibility have been developed [9, 10]. GOLD [11], GLIDE [12], Autodock Vina [13] and RosettaLigand [2] are some of the docking programs with receptor flexibility. GOLD allows side-chain receptor flexibility and ensemble docking [11]. In GLIDE, a protein structure prediction technique called Prime is used for treating protein flexibility [12]. Autodock Vina allows side-chain flexibility for the selected protein residues [13]. In RosettaLigand only side-chain conformational changes were initially allowed [2]. Recently, backbone flexibility has also been incorporated [14].

To address the problem of molecular docking, in our laboratory we have developed a docking program called MOLSDOCK [15, 16], which uses the mutually orthogonal Latin squares (MOLS) technique [17]. The MOLS sampling algorithm identifies a small sample of the vast multidimensional search space, which is nevertheless completely representative of this space. The energy values are calculated at each of the sampled points. A variant of the mean field method [18] analyses these energy values simultaneously to obtain the optimal conformation [19]. MOLSDOCK was tested for docking peptides [15], small molecules [16] and nucleotides [20]. In all the above cases, only the ligand was treated as flexible and the receptor was kept rigid. Later, we upgraded MOLSDOCK by incorporating receptor protein side-chain flexibility [21]. The upgraded version of MOLSDOCK, called iMOLSDOCK, was benchmarked and tested for docking peptides. However nonpeptide small organic molecules are most commonly used as drugs. Therefore we have further extended the 'induced-fit' docking in iMOLSDOCK to dock nonpeptide small molecule ligands, and in this paper we report this extension. We have benchmarked iMOLSDOCK with this extension using 34 protein-ligand complexes selected from the Protein Data Bank (PDB). We also have compared the performance of iMOLSDOCK with two popular flexible receptor docking tools : AutoDock Vina [13] and GOLD v5.2.1 [11]. We present these results here.

**Materials and Methods**

iMOLSDOCK is a 'flexible receptor/flexible peptide' docking method developed using the MOLS method [17]. The MOLS method has been described in detail elsewhere [17, 22, 23]. For completeness, we give a brief description here. We will first explain the method by applying it to the prediction of the minimum energy structure of a peptide. The MOLS method systematically searches the conformational energy space of the peptide to arrive at the optimum peptide structure. The conformational space of a peptide may be described as the set of all possible combinations of all values of all its variable torsion angles. Consider a peptide having '$m$' torsion angles, with each torsion angle taking up '$n$' different values. Then, there are $(n)^m$ conformations for the peptide. The potential energy for any of these conformations of the peptide may be calculated using an empirical energy function such as AMBER force field [24]. The next task is to locate the minimum energy conformation. Any attempt to search through all the possible $(n)^m$ conformations of the peptide will lead to combinatorial explosion. In MOLS method, the method of mutually orthogonal Latin squares [25] is used to systematically choose a set of $(n)^2$ points (or conformations) from the $(n)^m$ overall conformational space of the peptide. After choosing the $(n)^2$ conformations, the potential energy for each of these chosen conformations is calculated. The $(n)^2$ energy values are then analyzed using a variant of the mean field technique [18, 19] to arrive at the lowest energy conformation. The $(n)^2$ points could be selected in large number of ways [26], and each choice leads to either the same, or to a different low energy structure. We have shown repeatedly [15, 21, 27] that for small peptides of length up to about 10 residues, choosing the $(n)^2$ points in 1500 different ways (which would yield 1500 low energy conformations) is sufficient to identify all the unique, mutually dissimilar low energy structures.

The MOLS method was extended to address the docking problem [15, 16, 20]. The docking tool, dubbed MOLSDOCK, performed 'flexible ligand - rigid receptor' docking. In MOLSDOCK, the variables specifying the position and orientation of the ligand were added to the set of variable torsion angles of the ligand to define the search space. Accordingly, the scoring function for MOLSDOCK was also modified to include the interaction energy between the protein and the ligand, along with the intramolecular energy of the ligand. MOLSDOCK was developed as a program suitable for studying the docking of peptides, nonpeptide small organic molecules and nucleotides to protein receptors [15, 16, 20]. MOLSDOCK was upgraded to iMOLSDOCK [21] by including receptor flexibility. Two major changes were made in iMOLSDOCK. a) The search space was further increased to include the side-chain torsion angles of the receptor protein. b) The intra-protein energy, to assess the receptor protein conformation, was added along to the intra-ligand energy and protein-ligand interaction energy.

Thus, we specify the conformation of the small molecule ligand by '$s$' torsion angles ($\theta_r$, r = 1, $s$). Six additional parameters describe the ligand's docking pose, i.e. three for the position and three for the orientation of the ligand in the receptor binding site. If there are '$p$' torsion angles that describe the flexible residues in the receptor binding site, then we have a total of '$s+6+p$' dimensions in the search space ($\theta_r$, r = 1, $s+6+p$). The volume of the search space is $(n)^{s+6+p}$, if each dimension is sampled at '$n$' intervals. Out of the $(n)^{s+6+p}$ points in the search space, the MOLS method calculates the values of the scoring function only at $(n)^2$ points, and analyzes them to simultaneously locate the optimum conformation of the ligand, its pose, and also the conformation of the side-chains of the receptor flexible residues. Since the search space is defined on a discrete grid of '$s+6+p$' points, the actual optimum may be off-grid and lie close to but not actually on the grid point. Therefore to identify the nearest off-grid optimum, we perform a gradient minimization [28] as the final step.

The ligand (any organic chemical compound) is specified in MDL Molfile (.mol) format. The rotatable bonds in the ligand are identified using *findrotatable.pl* which is available at http://www.ccl.net/cca/software/PERL/Find_Rotatable_Bonds/. All the variable torsion angles of the ligand are sampled from 0º to 360º at intervals of 10º. As described by Arun Prasad and Gautham (2008), the orientation of the ligand in the binding site is specified by three angles, two of which specify the position of a rotation axis for the ligand. The third angle is the angle of rotation about this axis. The ligand translates inside a 5 Å cubic box centered at the midpoint of the receptor binding site. To test successful docking in iMOLSDOCK, it is necessary to dock the ligands to a pre-defined binding site before blind docking [29]. Therefore, for all the test cases (Table 1), we defined the binding site of the ligand. The binding site information was extracted from the holo protein-ligand crystal structure. (If the binding site in the receptor protein is not known, then binding site may be automatically found in iMOLSDOCK using the Fpocket 2.0 algorithm [30, 31]).

The 3-dimensional structure of the receptor protein is specified in PDB format. In iMOLSDOCK, side-chain receptor flexibility is allowed [21]. The flexible residues may be specified explicitly. In case, the flexible residues are not known, then residues that are within 4.0 Å from each atom of the ligand are automatically selected. In iMOLSDOCK, a maximum of 50 protein residues may be allowed to be flexible. For the test cases (Table 1), we allowed the side-chain torsion angles of the flexible residues to fluctuate in a range of 40° (i.e. -20° to +20° from their position in the crystal structure).

The scoring function in iMOLSDOCK is the weighted sum of intra-ligand energy, protein-ligand interaction energy and intra-protein energy. The intra-protein energy is calculated using the AMBER94 force field [24]. The intermolecular interaction energy is calculated using the PLP scoring function [32]. MMFF94 force field works efficiently for small molecule ligands [33]. General AMBER force field (GAFF) is a general force field where all the parameters are available and covers almost all the organic chemical spaces [34]. In iMOLSDOCK, the intra-ligand energy may be calculated either using MMFF94 force field [35, 36] or GAFF [34]. We have tried both MMFF94 and GAFF as intra-ligand energy separately for our test cases. As stated earlier (Paul and Gautham, 2017), we use the PLP force field [32] for the protein-ligand interaction energy, and AMBER force field [24] for the intra-protein energy. The total energy is a weighted sum of these three terms. To fix the weights for each term, calculations were performed on a small subset of the structures. The optimum weights were chosen to yield maximum positive correlation between the energy and the root mean square deviation (RMSD) of the resulting docked structure with respect to the 'native' crystal structure. The AMBER force field [24], MMFF94 [35, 36] and GAFF [34] are expressed in units of kcal/mol, whereas the PLP force field is reported in dimensionless units [32]. Therefore the total potential energy is also reported in dimensionless units.

iMOLSDOCK is a command-line-only FORTRAN-based induced-fit protein-ligand docking tool. To make iMOLSDOCK easily assessable to the scientific community, we have developed a Java-based Graphical User Interface (GUI) and added it to MOLS 2.0 [31]. MOLS 2.0 is a software package developed in our laboratory. The software package is available free at https://sourceforge.net/projects/mols2-0/files/.

We have used a set of 34 protein-ligand complexes for which the crystal structures of both *apo* and *holo* forms were available, taken from the PDB, to test the performance of iMOLSDOCK (Table 1). The binding sites in these proteins are not similar to each other, and the proteins belong to different families of the SCOP database [37]. The number of variable torsion angles in the ligands varies from 1 to 19. For all the test cases, the ligands were docked into the *apo* form of the receptor protein structure. Details of the *apo* and *holo* form of the receptor protein for all test cases are given in Table 1. We carried out induced-fit docking for all the cases, i.e. while docking in iMOLSDOCK, the conformation of the ligand, as well as conformation of the flexible residues lining the binding site in the receptor protein alter simultaneously.

**Results and discussion**

A total of 1,500 structures were generated for all the 34 cases of the benchmarking dataset. Throughout the analysis, the crystal structure was considered as the native structure. In each case, out of the total 1,500 structures, two structures were selected for our analysis. The first structure, called the 'best sampled' structure, is the prediction that has the lowest RMSD with respect to the native structure. The best sampled structure is found by superimposing all the 1,500 predicted protein-ligand complex structures on the respective native protein-ligand complex structure by least squares superposition of all the atoms, without altering either the structure of the ligand, or its position and orientation relative to the protein. The RMSD was then calculated on all the heavy (non-hydrogen) atoms of the ligand. This particular method measures the differences in the docking pose of the ligand with respect to the protein along with the difference in the ligand structure. The second structure we used in our analysis is the prediction with the lowest energy of the total 1,500 predictions. This is the energetically top-ranked structure.

Results obtained with GAFF and MMFF94 as intra-ligand energy function are summarized in Supplementary Table 1. The RMSD and energy of the top-ranked structure and the best sampled structure are given in this Table. In the results with GAFF as the intra-ligand energy function (hereafter, GAFF-results), the RMSD of the best sampled structure is less than or equal to 2.50 Å in 16 of the 34 cases. In the results with MMFF94 force field as the intra-ligand energy function (hereafter, MMFF-results), the RMSD of the best sampled structures is less than or equal to 2.50 Å in 19 of the 34 cases. A brief summary of the overall results is given in Table 2. The best sampled structure of GAFF-results for all the cases is shown in Supplementary Figure 1 and the best sampled structure of MMFF-results for all the cases is shown in Supplementary Figure 2.

*Ranking efficiency*

An ideal scoring function in a docking tool is expected to top-score the best sampled (near-native) structure. A solution for which the best sampled structure is the same as the top-ranked structure is defined as an exact solution. In GAFF-results, exact solution was found for case 2YPI. In both GAFF-results and MMFF-results, for 7 of the 34 cases the best sampled structure is among the top 10% energy-ranked solutions. The top-ranked solution of 1NGP and 2PK4 have RMSD ≤ 2.00 Å from the native structure in GAFF-results. In MMFF-results, the top-ranked solutions of 1TNK, 2PK4 and 1AJQ have RMSD ≤ 2.00 Å from the native structure. It has been observed that in 8 of the 34 cases, in both GAFF-results and MMFF-results, at least one structure with RMSD ≤ 2.00 Å is among the top 10% ranked solutions.

*Hydrogen bond interactions*

The hydrogen bond interactions between the protein and the ligand were found using HBPLUS [38]. A comparison of these interactions in the native complex and in the top-ranked complex determined by iMOLSDOCK is shown in Figure 1. Hydrogen bond prediction in GAFF-results is better than MMFF-results for 25 of the 34 cases. In 10 of the 34 cases, the top-rank structure predicted using GAFF has more hydrogen bond interactions than observed in the native structure. In 6 cases using MMFF94 resulted in more hydrogen bonds than the native structure. For the cases 4EST and 1BLH, no hydrogen bond interactions were predicted by MMFF94 in the top-ranked structure, while GAFF predicted 2 hydrogen bonds in the top-ranked structure of 4EST and 4 hydrogen bonds in the top-ranked structure of 1BLH respectively. In many cases the hydrogen bond interactions seen in the native structure are also predicted by iMOLSDOCK in the top-ranked solutions. In cases 1NGP and 2PRH, all the hydrogen bond interactions predicted by MMFF94 as the top-ranked solution are also seen in the native crystal structure (Figure 2). The RMSD of the top-ranked solution of 1NGP from the native structure is 2.08 Å. The RMSD of the top-ranked structure of 2PRH from the native structure is 5.37 Å.

*Alternate binding modes*

Alternate solutions that have a lower energy value than the native structure are often detected by algorithms [39]. iMOLSDOCK does not converge to just a single solution but generates hundreds of low-energy possibilities. In certain cases, the native structure, the best sampled structure and the lowest energy structure (which is the top-ranked structure) are iso-energetic, i.e. they are energetically indistinguishable [40]. In GAFF-results, for 6 of the 34 test cases, the lowest energy structure shows an equally favourable alternate binding mode (Figure 3(a)). In MMFF-results, for 9 of the 34 test cases, the lowest energy structure shows an equally favourable alternate binding mode (Figure 3(b)). These alternate binding modes are characterized by energy values lower than native energy, but with larger values of RMSD as compared to the native complex. Alternate binding modes are important in the study of the possible toxicity of putative drugs [41].

*Comparison studies between Vina, GOLD and iMOLSDOCK*

We chose Autodock Vina [13] – a free and open-source docking tool – and GOLD [11] – a commercial and widely used docking tool – for the comparison. The key aspects that are taken into consideration during the comparison are sampling efficiency, ranking efficiency, prediction of non-bonded interactions and computation time of iMOLSDOCK, Vina and GOLD. The iMOLSDOCK results obtained with General Amber Force Field [34] as the intra-ligand energy are considered for the comparison studies.

We carried out small molecule docking in Vina for all the 34 protein-ligand complexes (Table 1). The input structure of ligand and protein were prepared using AutoDock Tools (ADT; [42]. The input structure of the ligand was taken from the bound complex crystal available in the PDB. We used ADT to randomize the starting conformation, position and orientation of the ligand. The apo protein structure taken from the PDB (see Table 1) was used as the receptor protein structure for all the cases. The binding site was predefined for all the test cases. On an average, 10 residues lining the binding site in the receptor protein were selected to be flexible. We set the search space to 22.5 Å, the maximum dimension suggested by the authors [13], in all the three axes. The other parameters (exhaustiveness, number of modes, etc.) were fixed to the default values.

The crystal structure of the ligand was taken from the protein-ligand complex and was used as the input ligand structure. We defined the binding site for all the cases. To allow receptor flexibility, we allowed 10 flexible residues in the binding site to be flexible. GOLD allows partial protein flexibility in the neighbourhood of the protein active site. We chose the 'Crystal' option under 'Rotamer Library Operations' for protein side-chain flexibility. GoldScore was selected as the fitness function for all the cases. GoldScore is formulated in such a way that, the larger the score, the better the docking result is likely to be. GoldScore is dimensionless. Throughout the small molecule docking, the parameters of the fitness function were fixed to their default values. A total of 150 GA runs were performed for each test case. All the other operations were set to 'automatic'.

*Sampling efficiency*

Supplementary Table 2 shows a comparison of the results obtained from the three programs. Out of all the docked structures for each test case, we chose the best sampled structure and the top-ranked structure for our analyses. The best sampled structure is the structure with the lowest RMSD with respect to the native complex. The lowest energy structure is the top-ranked structure.

With Vina, the best sampled solutions of all the cases except 1TNH are 2.00 Å or more away from the crystal structure. With Gold, in 11 of the 34 cases, the best sampled structures are within 2.00 Å from the native structure. With iMOLSDOCK, in 10 of the 34 cases, the best sampled structures are within 2.00 Å from the native structure

With Vina, the top-ranked structures of 1TNH is within 2.00 Å from the crystal structure. With GOLD the top-ranked solution in 4 cases (1TPP, 1NGP, 1AEC, and 1AJQ) are within the RMSD of 2.00 Å from the native structure. With iMOLSDOCK the top-ranked solution in 2 cases (1NGP and 2PK4) are within the RMSD of 2.00 Å from the native structure.

Thus, both in terms of the best sampled solutions, as well as the top-ranked structures, iMOLSDOCK performs better that the other two algorithms.

*Scoring efficiency*

With Vina, we generated the default number of binding modes, 9 in each of the 34 cases. Vina is able to identify exact solutions – the lowest RMSD structure is also the top-ranked structure – in 4 cases (2PRH, 2YPI, 2XIS, and 1AAQ). GOLD predicted exact solutions in 7 cases (1NGP, 4EST, 1PDZ, 1BLH, 1AEC, 1AI4, and 1AJQ). Of these, 3 cases are within 2.00 Å from the native structure. iMOLSDOCK identified exact solution in one case (2YPI).

With Vina, except for the exact solutions, none of the cases have the best sampled structure among the top 10% energy-ranked solutions. With GOLD, in 27 out of the 34 cases the best sampled structure is among the top 10% of the ranked structures. With iMOLSDOCK, the best sampled structures in 8 of the 34 cases are among the top 10% of the energy-ranked solutions.

Clearly GOLD performs the best in this measure, but iMOLSDOCK results are comparable to those obtained with Vina.

*Hydrogen bond prediction*

The number of hydrogen bond interactions in the top-ranked structures of Vina and GOLD and iMOLSDOCK are shown in Figure 4. iMOLSDOCK predicted more hydrogen bonds in the top-ranked structures than Vina or GOLD in 15 and 11, respectively, of the 34 cases. In 5 cases (1TNH, 1NGP, 1TPH, 1CBX, and 1AI7), the number of hydrogen bonds in the top-ranked structure are the same in both iMOLSDOCK and Vina. In 7 cases (1TPP, 3MTH, 4EST, 1EBG, 2PRH, 2YPI, and 1BLH), the number of hydrogen bonds in the top-ranked structure are equal in both iMOLSDOCK and GOLD. Overall, GOLD predicts a larger number of hydrogen bonds for each structure than iMOLSDOCK.

*Computation time comparison between iMOLSDOCK, Vina and GOLD*

The computation time of Vina, GOLD, and iMOLSDOCK with respect to the number of torsion angles (ligand torsion angles + protein torsion angles) are shown in Figure 5 (See also Supplementary Table 3). The average computation time for one run in Vina is 3.1 m. In GOLD it is 1.8 m. Both these values are much smaller than for iMOLSDOCK (1.3 h per structure). iMOLSDOCK however handles more variable parameters (torsion angles) than Vina or GOLD. In our test cases, we allowed 20 flexible residues in iMOLSDOCK, 10 flexible residues in Vina and in GOLD.

*iMOLSDOCK performance against the Astex Diverse set of protein-ligand complex*

Of the 85 cases of the Astex Diverse set [43], we selected 55 protein-ligand complexes that had no ligand - metal ion interaction. We tested the performance of iMOLSDOCK for 55 cases of the Astex Diverse set taken from the Cambridge Crystallographic Data Centre and then compared the performance of iMOLSDOCK with other docking tools: GOLD [11] and Autodock Vina [13]. The overall comparison results are shown in Supplementary Table 4. For all the 55 cases, 1500 models were generated as discussed in the methods section. In 27 of the 55 cases the best sampled structure has an RMSD lesser than 2.00 Å. In 10 of the 55 cases, the best sampled structure has an RMSD lesser than 1.00 Å. For one case (1s3v), iMOLSDOCK predicted the best sampled structure as the lowest energy structure i.e., the top-ranked structure. However, for 30 of the 55 cases the best sampled solution is within the top 10% when ranked in terms of energy. The top-ranked structure is within 2.00 Å from the native structure in 13 of the 55 cases. Of the 55 cases, GOLD predicted exact solutions i.e., the best sampled structure is the top-ranked structure, for 32 cases and Vina predicted exact solutions for 38 cases respectively. Of the 55 cases, the best sampled structure predicted by GOLD and Vina is within 2.00 Å for 40 and 43 cases respectively.

*Cross-docking*

Cross-docking experiment is carried out in situations where a crystal structure of the protein with a ligand bound in the binding site is known. During drug discovery, the binding of different ligands to the same protein needs to be evaluated accounting for potential changes in the protein side-chain or backbone conformation [2]. Firstly, for our cross-docking experiment, we randomly selected the crystal structure of *Penicillin acylase enzyme* bound with ten different ligands from the Astex non-native dataset [44]. The ligand was cross-docked to the non-native conformer of the same protein using iMOLSDOCK. The results are shown in Supplementary Table 5. For all the 10 cases, the best sampled structure of the ligand is within 2.01 Å from the native structure. Secondly, for our cross-docking experiment, we randomly selected 10 cases from the SEQ17 dataset [45] which contains apo and holo pairs for a diverse set of receptors [45]. This cross-docking experiment is carried to check whether iMOLSDOCK is able to perform the protein side-chain conformational change that is required for ligand binding as rigid docking is likely to fail for these cases. The results obtained were compared with the results obtained from GOLD docking tool. RMSD of the best sampled solution of iMOLSDOCK and GOLD are shown in Supplementary Table 6. Except for one case (1it8), iMOLSDOCK was not able to bind the ligand at least within 2.00 Å from the native structure. In case 1it8, the ligand could approach the active site, as iMOLSDOCK modified the side-chain torsion angle of the active site residue Phe229 from 99.4° (apo form which is the starting structure) to 60.0° (native = 68.8°) thereby enabling the ligand to bind with RMSD = 1.43 Å from the native ligand. However, for case 1k4h, iMOLSDOCK was not able to alter the conformation of Tyr106, the active site residue blocking the entry of the ligand into the active site. Whereas GOLD, using the 'rotamer' option for receptor flexibility, was able to give a rotameric shift to Tyr106 enabling ligand binding with RMSD = 0.85 Å from the native structure. Side-chain torsion angle difference between the apo and the holo form of the active site residue Tyr106 is 109.2°. iMOLSDOCK is currently equipped to provide small side-chain fluctuations to the flexible residues in the receptor active site.

The aim of the current paper is to improve iMOLSDOCK by enabling small molecular – protein docking with induced-fit receptor flexibility. The docking failures observed has shown us areas where iMOLSDOCK and the scoring function may be improved. Especially, the addition of rotameric side-chain flexibility along with ring-flipping [11, 46] algorithm might enhance iMOLSDOCK especially for cross-docking experiments.

**Conclusion**

We have extended the 'induced-fit' docking feature in iMOLSDOCK, from docking peptide ligands to dock drug-like small molecule ligands. Test runs using 34 protein-ligand complexes from the PDB show that the method performs well. We have tested iMOLSDOCK with two different force fields (GAFF and MMFF94) for calculating the intra-ligand energy of the scoring function. Comparison of iMOLSDOCK with Autodock Vina and GOLD shows that sampling in iMOLSDOCK is better than the other two tools. The search space in iMOLSDOCK is much greater than in Vina or GOLD as it allows for more flexible residues than Vina and GOLD. iMOLSDOCK is also able to predict alternate binding modes.